# From Prompts to Constructs: A Dual-Validity Framework for Large Language Model Research in Psychology

Zhicheng Lin

Department of Psychology, Yonsei University, Seoul, South Korea; email: zhichenglin@gmail.com

**Abstract**

Large language models (LLMs) are entering psychological research both as tools and as objects of inquiry. Yet many studies apply human instruments to LLMs without establishing that the outputs are reliable or interpretable, raising the risk of measurement phantoms—statistical regularities mistaken for genuine psychological phenomena. This review argues that robust AI psychological research requires integrating two methodological traditions: psychometric validation of what a score means and causal inference standards for what the results warrant. It develops a dual-validity framework in which evidentiary demands scale with scientific ambition: from tool use through behavioral characterization and human simulation to cognitive modeling. Classifying text may require only accuracy and reliability; claiming that an LLM simulates anxiety or illuminates cognitive mechanisms requires additional evidence, including construct validity evidence and experimental controls. Progress depends on developing computational analogs of psychological constructs rather than assuming human measures automatically apply to language models.

## INTRODUCTION

When researchers tested GPT models on moral dilemma scenarios, they reported human-like ethical preferences: The models seemed to value saving more lives and preserving humans over animals (Takemoto 2024). However, Oh & Demberg (2025) discovered something troubling: Simply changing “Case 1” and “Case 2” to “(A)” and “(B)” reversed many of these moral preferences; recasting “Respond with one word only.” as “Can you respond with one word only?” also altered judgments. Moral reasoning in these models was as sensitive to punctuation as to ethical principles.

These vulnerabilities extend beyond moral reasoning and point to methodological challenges in large language model (LLM) psychological research (Löhn et al. 2024, Schelb et al. 2026, Shapira et al. 2024, Voudouris et al. 2025, Ye et al. 2025a). By repurposing psychological inventories, questionnaires, and behavioral tasks originally developed for human participants, studies now claim to measure personality traits, theory of mind, cognitive biases, and emotional intelligence in language models, often drawing direct parallels to human psychology (Binz & Schulz 2023, Kosinski 2024, Miotto et al. 2022, Pellert et al. 2024, Wang et al. 2023, Webb et al. 2023). This practice rests on untested assumptions: that human-validated instruments remain valid when administered to nonhuman systems, that surface-similar outputs reflect similar underlying processes, and that statistical patterns in model responses carry the same inferential weight as patterns in human data (for a related taxonomy of interpretive fallacies, see Lin 2025a).

Each assumption is empirically testable (Cronbach & Meehl 1955), and emerging evidence suggests that model responses violate the psychometric conditions that these assumptions require (Gao et al. 2025, Lee et al. 2025, Peereboom et al. 2025, Tjuatja et al. 2024, Wang et al. 2025b). For example, trivial prompt perturbations—such as adding extra spaces, altering punctuation, or changing the order of few-shot examples—can produce variation of up to 76 percentage points in accuracy across prompt formats (Guan et al. 2025, He et al. 2024, Sclar et al. 2024). Models may simultaneously agree with contradictory items—both “I am an extrovert” and “I am an introvert” (Sühr et al. 2025).

If moral preferences flip with parentheses, can we trust the measurement itself? And if we cannot reliably measure moral reasoning, how can we test whether experimental manipulations—different scenarios, cultural contexts, or prompt structures—causally affect it? Unreliable measurement thus cascades through experimental design, undermining both construct validity and causal claims. Some studies treat LLM responses as windows into psychological processes, interpreting behavioral patterns as evidence of underlying cognitive mechanisms (Sartori & Orrù 2023). Others acknowledge that LLMs simulate responses but draw causal conclusions without addressing either computational confounds arising from technical artifacts or the limits of observational designs (e.g., Binz & Schulz 2023, Dillion et al. 2023, Kosinski 2024, Park et al. 2023).

> **Construct validity:** accumulated evidence supporting a particular interpretation of scores as indicators of a theoretical construct

These pervasive problems boil down to two interrelated questions. First, can LLM responses measure psychological constructs reliably, as indexed by internal consistency, test–retest reliability, and parallel-forms reliability? Second, when reliable measurement exists, what types of scientific inferences can we draw? These inferences include both descriptive claims about LLM properties (does this model exhibit trait X?) and causal claims about experimental manipulations (does intervention Y affect behavior Z?).

> **Internal consistency:** degree to which multiple indicators of one construct yield coherent, intercorrelated responses

**Test–retest reliability:** stability of scores when the same instrument is readministered to the same entity on separate occasions

**Parallel-forms reliability:** consistency of scores across semantically equivalent prompt versions designed to measure the same construct

Answering these questions requires integrating two validity traditions that have evolved separately in psychological methodology: the psychometric tradition, which asks whether instruments measure the intended constructs (Cronbach & Meehl 1955, Loevinger 1957, Messick 1989), and the causal inference tradition, which asks whether studies support valid conclusions about cause and effect (Campbell & Stanley 1963, Cook & Campbell 1979). While human research typically emphasizes one or the other, LLM research demands both: establishing that prompts and responses constitute valid measures, then ensuring that research designs support appropriate inferences (see the sidebar titled Two Validity Traditions).

### SIDEBAR 1: TWO VALIDITY TRADITIONS

The psychometric tradition, rooted in educational and psychological testing, asks what a score means: Can prompts and outputs be interpreted as indicators of a construct? Evidence accumulates across five sources: content, response processes, internal structure, relations to other variables, and consequences of testing.

The causal inference tradition, developed for experimental and quasi-experimental research, asks what a study warrants: Can an observed change be attributed to a manipulation and generalized appropriately? Evidence is evaluated against threats to four parallel validity types: internal validity, external validity, construct validity of causes and effects, and statistical conclusion validity.

LLM psychology requires both traditions because, when the model is itself the object of inquiry—whether for behavioral characterization, human simulation, or cognitive modeling—the same output often serves as both a measurement indicator and an experimental outcome. If the measure is unreliable or lacks construct validity, causal inference is built on noise. If the design is confounded, apparent construct evidence may reflect artifacts of prompts, parameters, or model versions rather than the construct itself. In tool-use applications, by contrast, the measurement target lies outside the model, and the two roles separate.

This integration forms the foundation for understanding validity in LLM psychological research. I first establish how the psychometric and causal inference traditions apply to LLM research contexts; then I examine how reliability failures undermine both measurement and inference foundations. Finally, I show how construct validity evidence must be accumulated across multiple sources and how experimental designs must address threats to four parallel validity types in causal inference. Because the field is young and many of these validity issues have only recently come into focus, the aim is not to assign retrospective blame but to specify, for each level of scientific ambition, what evidence is needed before a claim can bear its own weight.

## TWO VALIDITY TRADITIONS AND THEIR INTEGRATION IN LLM RESEARCH

### The Psychometric Tradition

In the psychometric tradition, the central problem is measurement quality (Cronbach & Meehl 1955, Loevinger 1957). Researchers need to know whether intelligence tests, personality inventories, and attitude scales capture their intended constructs. Validity is thus about meaning: What do scores signify? This question precedes all others; without valid measurement, subsequent analyses become exercises in quantifying noise.

In this tradition, validity is treated as a single overarching interpretive argument—construct validity—to which multiple kinds of evidence contribute (Messick 1989). (This usage differs from the causal inference tradition, where construct validity is one of several parallel validity types.) Evidence flows from five principal sources: content (do items sample the construct domain?), response processes (do test takers engage in the expected cognitive operations?), internal structure (do responses show theoretically consistent patterns?), relations to other variables (do scores correlate as theory predicts?), and consequences (what are the implications of score interpretations?).

> **Relations to other variables:** evidence that scores show expected convergent, discriminant, and predictive associations with external criteria

However, for a score interpretation to be meaningful, the instrument must be sensitive to variations in the target attribute—a link traditionally investigated through response process evidence. A causal theory of validity makes this requirement explicit: A measure is valid only if (a) the attribute it is supposed to measure is genuinely present in the target system, and (b) changes in that attribute causally produce changes in the observed score (Borsboom et al. 2004). This point is especially consequential in LLM research, because favorable psychometric statistics (e.g., Li et al. 2025b) do not answer the prior question of what, if anything, the score refers to in the model. A model might endorse "I worry about the future," but anxiety in humans presupposes temporal experience, a persistent self, and embodied consequences of threat—features not established for current language models (Lin 2025a). The problem is therefore not only whether outputs show a pattern, but what that pattern is evidence of.

> **Response process evidence:** evidence that the mechanisms generating outputs align with the processes that the intended score interpretation presupposes

Building on Messick's (1989) unified treatment of construct validity (Loevinger 1957), the modern validity framework, codified in the Standards for Educational and Psychological Testing (American Educational Research Association et al. 2014), establishes that validity is not a property of an instrument in the abstract but an argument—grounded in accumulated evidence—for a particular interpretation of its scores. A scale validated for humans does not remain valid simply because a language model can produce plausible answers to it; changing the respondent changes what the score could mean. For that reason, human psychological attributes cannot be taken as a starting assumption in LLM research; they must be justified as part of the validation argument itself. This reframing points toward computational analogs of psychological constructs—reconceptualizations that preserve a construct's theoretical role while respecting mechanistic differences between biological and computational systems.

> **Computational analog:** a reconceptualization of a psychological construct that preserves its theoretical role while acknowledging mechanistic differences between biological and computational systems

### The Causal Inference Tradition

In the causal inference tradition, the central problem is drawing warranted conclusions about cause and effect (Campbell & Stanley 1963, Cook & Campbell 1979). Researchers need to know whether the observed effects can be causally attributed to treatments, interventions, and manipulations. Validity is thus about warranted inference: Can we trust our conclusions? The framework assumes meaningful measurement exists and focuses on threats to causal reasoning.

Rather than hierarchical evidence accumulation, this approach conceptualizes validity through four parallel types, each addressing distinct threats to causal inference: Internal validity asks whether observed changes stem from manipulations rather than confounds; external validity examines whether causal relationships generalize beyond specific studies; construct validity of causes and effects evaluates whether operational definitions align with theoretical constructs; and statistical conclusion validity addresses whether data analyses support causal inferences. These validity types are parallel, not hierarchical: A study might demonstrate strong internal validity (the manipulation caused the change) while suffering from weak external validity (the effect does not generalize) or construct validity problems (the manipulation does not represent the intended theoretical variable).

### Why LLM Research Requires Both

To understand why LLM research demands integrating both validity traditions, it is necessary to distinguish two broad roles that LLMs play in psychological research. In the first, the LLM is a research tool—to code qualitative data, extract sentiment, generate stimuli, or assist measurement (Binz et al. 2025; Blanchard et al. 2025; Demszky et al. 2023; Feuerriegel et al. 2025; Lin 2023, 2025b; Ziems et al. 2024). The inferential target lies outside the model: Researchers care about the text being coded, the construct being scored, or the stimulus being produced.

In the second role, the LLM is itself the object of inquiry. Here, three claim types should be separated by inferential target: Behavioral characterization (machine psychology, or GPT-ology) asks what patterns the model shows under specified prompts and tasks (Li et al. 2024, Ong 2024, Sartori & Orrù 2023); human simulation asks how closely those patterns match human data from a specified population or experiment (Dillion et al. 2023, Grossmann et al. 2023, Lin 2026b); and cognitive modeling asks whether the processes producing those outputs illuminate human cognitive mechanisms (Blank 2023, Frank & Goodman 2026, Lin 2026b, Niu et al. 2026).

This hierarchy clarifies the validity burden: Stronger claims build on the validation requirements of weaker ones (Figure 1). Tool use is primarily a problem of task validation: accuracy against a given criterion, robustness to prompt variation, and calibration of probabilistic outputs. Building a psychological instrument from model outputs additionally requires internal consistency and selective external correlations (Schlegel et al. 2025, Xu et al. 2024). Once the LLM becomes the object of psychological inquiry, however, construct validity is central from the outset—and the burden differs across the three subtypes.

Behavioral characterization supports descriptive claims about the model and therefore requires construct-sensitive psychometric evidence plus causal validation of testing conditions before psychological labels are warranted (Millière & Buckner 2024). When researchers report that LLMs exhibit theory of mind (Kosinski 2024), display personality traits (Miotto et al. 2022), or demonstrate moral reasoning (Takemoto 2024), they make measurement assertions that demand psychometric validation.

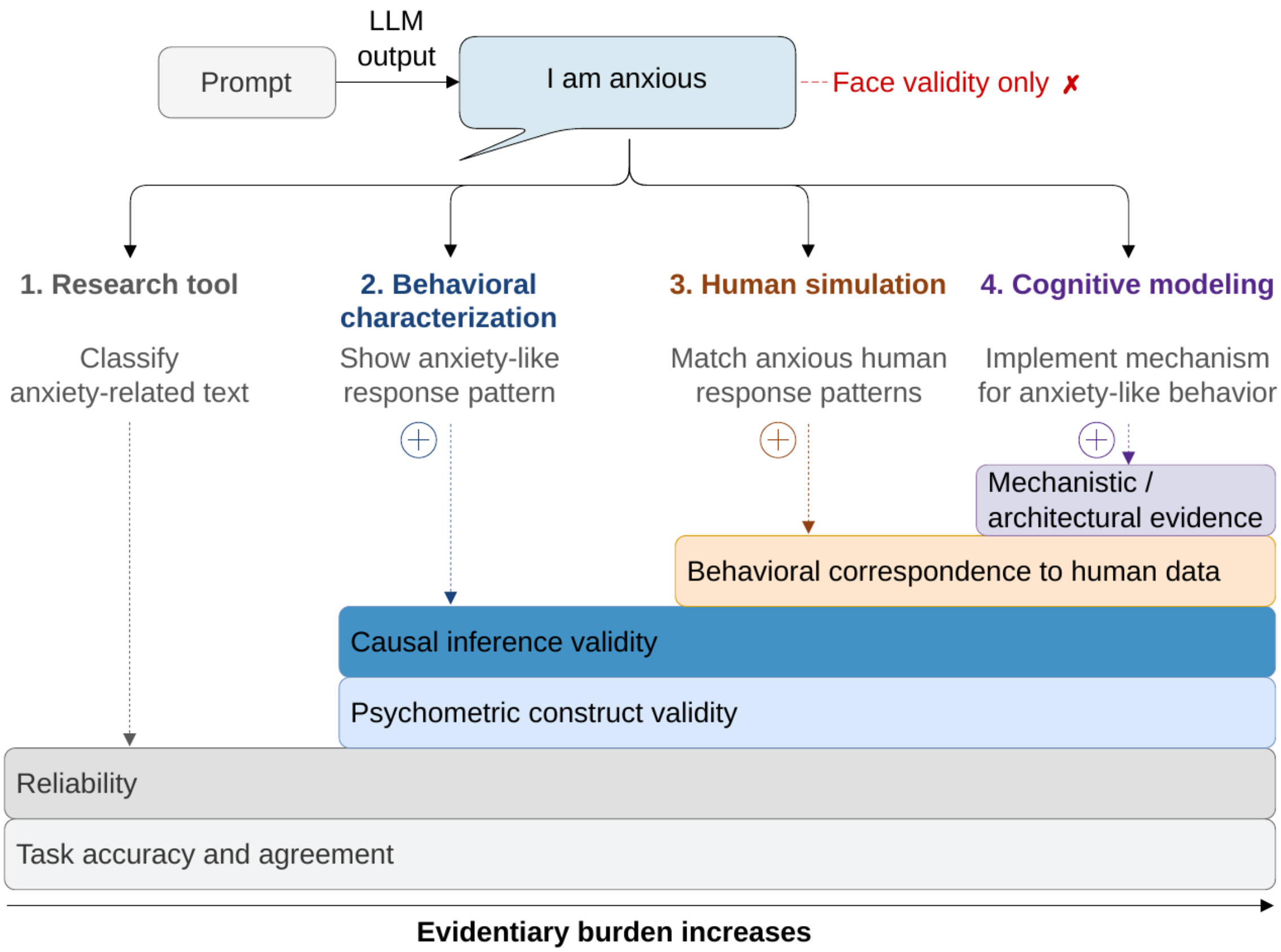

**Figure 1** Scientific ambition determines evidentiary burden in large language model (LLM) psychology. The same LLM output—here, "I am anxious"—can serve either instrumentally, as a research tool, or substantively, as the object of psychological inquiry. Within the latter role, requirements accumulate from behavioral characterization through human simulation to cognitive modeling, as indicated by the dotted lines with plus symbols. ✗ marks the face-validity shortcut: treating surface plausibility as sufficient

evidence for psychological claims without the cumulative validation shown below. Such shortcuts may produce measurement phantoms.

Human simulation adds a fidelity requirement: The model must reproduce relevant human response distributions or experimental effects, and that correspondence must survive construct-preserving variation in prompt format, task design, and benchmark population (Lin 2026b). Cognitive modeling imposes the strongest burden: evidence that the model's computational processes or internal representations are informative about human cognition, not just that its outputs resemble human answers (Guest & Martin 2023, Lin 2026b). Causal requirements escalate in parallel: Descriptive tool use may involve no causal claims, but studies that manipulate prompts, scenarios, or contexts to examine cultural differences, ethical preferences, or cognitive biases (Grossmann et al. 2023) advance causal hypotheses, requiring safeguards against confounds, generalization failures, construct misalignment, and statistical artifacts.

The two validity traditions converge when the same LLM output is treated both as a measurement indicator and as an experimental outcome. A model generates text about moral dilemmas, and researchers interpret that text as evidence about a psychological construct despite the known psychometric fragility of such measures (Gao et al. 2025, Peereboom et al. 2025). At the same time, they treat variation in that text as evidence that prompts or scenarios have produced construct-relevant effects, despite uncontrolled confounding introduced by prompt design and model architecture (Bisbee et al. 2024, Gui & Toubia 2023). The result is a cascading vulnerability: Weak measurement undermines descriptive claims about the model, surrogate claims about humans, and mechanistic claims about cognition alike, while experimental confounds make apparent construct evidence difficult to interpret.

Yet many psychological studies using LLMs do not explicitly integrate either tradition (Demszky et al. 2023, Ivanova 2025, Löhn et al. 2024). Research claiming to measure models' personality, theory of mind, or moral reasoning proceeds without establishing measurement foundations (Peereboom et al. 2025, Wang et al. 2025b, Ying et al. 2025). Studies manipulating prompts to test psychological hypotheses often lack experimental safeguards, relying instead on face validity (the shortcut shown in **Figure 1**). This gap widens at the highest levels of the hierarchy, where validity demands are greatest and evidential shortfalls most consequential. **Table 1** maps each validity threat to its computational mechanism and to an example scenario, providing a diagnostic checklist for identifying these vulnerabilities.

> **Face validity:** the informal judgment that a measure appears to assess the intended construct based on surface inspection, without systematic psychometric evidence

**Table 1** Diagnostic map of validity threats in LLM psychological research

| Validity type | Threat | Mechanism | Example |
|---|---|---|---|
| **Psychometric tradition** | | | |

| Validity type | Threat | Mechanism | Example |
| --- | --- | --- | --- |
| Content | Domain undersampling | Prompts or items sample only a narrow, prototypical slice of the construct domain. | A moral reasoning study uses only trolley problem variants, ignoring care-based, virtue-based, or culturally specific moral domains. |
| Content | Training-data/item contamination | Items, close paraphrases, or benchmark formats are present in training data, eliciting recognition-based responding rather than construct-relevant measurement. | Big Five Inventory items appear verbatim in web-scraped corpora, so outputs reflect memorized associations rather than latent trait structure. |
| Response processes | Mechanistic substitution | Outputs are generated by next-token prediction rather than the psychological process the intended score interpretation presupposes. | A model endorses “I feel anxious in social situations” via statistical co-occurrence rather than a process resembling experiential anxiety. |
| Response processes | Architectural artifacts | Attention patterns, positional biases, or training objectives produce systematic response tendencies unrelated to the target construct. | Primacy and recency effects in multiple-choice items are driven by token-position attention weights, not by item content. |
| Internal structure | Structural misfit | Factor structures extracted from LLM responses fail to replicate the dimensionality established in human populations. | An analysis of Big Five Inventory responses yields three or seven factors instead of five, indicating that the covariance structure of the model’s responses differs from that of human responses. |
| Internal structure | Factorial collapse | Alignment-induced agreeableness or social-desirability bias compresses variance and inflates inter-item correlations, producing artificially high alpha coefficients. | Near-ceiling endorsement of socially desirable items yields a single general factor rather than differentiated trait dimensions. |

| Validity type | Threat | Mechanism | Example |
|---|---|---|---|
| Relations to other variables | Nomological instability | Convergent and discriminant correlations shift across models, prompt framings, measurement formats, or temperature settings, preventing stable network interpretation. | Self-report personality scores show mostly weak correlations with user-perceived personality and interaction quality in task-based dialogue, but align with assigned profiles and correlate to varying degrees with human raters' perceived-trait scores in a creative-story task. |
| Relations to other variables | Behavioral–report disconnect | Self-report-style endorsements do not predict model behavior on tasks theoretically linked to the same construct. | A model scores high on conscientiousness items but shows no corresponding pattern in task completion accuracy or deadline adherence simulations. |
| Consequential | Bias reification | Training-data biases (gender, racial, cultural stereotypes) are reproduced in model outputs and then interpreted as stable model traits or values. | Responses to some demographic prompts receive systematically lower competence ratings; researchers interpret this as the model's implicit attitude rather than a corpus artifact. |
| Consequential | Misguided application | Score interpretations are applied in contexts beyond the supporting validation argument, producing downstream harm. | Personality scores from a general-purpose chat model are used to characterize the psychological profiles of demographic groups the model was prompted to simulate. |
| **Causal inference tradition** | | | |
| Internal | Sampling-parameter confounding | Temperature, nucleus sampling (top-p probability cutoff), or decoding constraints covary with the intended experimental manipulation. | A framing-effect study compares conditions run at different temperature settings; a personality profile stable at 0.2 fragments at 1.0. |
| Internal | Prompt confounding | Experimental conditions differ in wording, syntax, language, length, system prompt content, or information position within the context window. | A cultural comparison confounds cultural identity with prompt language (e.g., Chinese versus English); conditions also differ in prompt length, so output differences may reflect verbosity rather |

| Validity type | Threat | Mechanism | Example |
|---|---|---|---|
| | | | than the treatment. |
| Internal | Version-drift confounding | Hidden model or API updates alter the system between data-collection waves, introducing uncontrolled variation. | A pre–post study spans a silent model update, so apparent treatment effects may reflect version differences rather than the experimental manipulation. |
| Internal | Background-reconstruction confounding | The model infers unstated background covariates from the treatment cue; attempts to fix covariates explicitly can make the specified information artificially salient. | Raising one variable leads the model to infer that correlated background variables have also shifted, distorting the causal estimate. |
| External | Generalization failure | Causal effects observed with one model, prompt set, or parameter configuration do not replicate with others. | An anchoring effect demonstrated in GPT-4 at temperature 0.7 disappears in Claude, LLaMA, or the same model at temperature 0. |
| External | Population mismatch | Effects observed in model outputs are projected onto human populations without validation against human data. | A simulation study claims that people show a particular decision bias based solely on LLM outputs, without any human behavioral data. |
| Construct | Manipulation-validity failure | The operationalized cause shifts text associations or token probabilities rather than instantiating the intended psychological mechanism. | A question like “Everyone agrees with X. What do you think?” changes agreement-associated token probabilities, not conformity processes involving status or belonging needs. |
| Construct | Outcome-validity mismatch | The dependent variable indexes surface-level language production rather than the psychological capacity the causal claim targets, so the causal effect operates on a different construct. | An emotional awareness scale captures the model’s linguistic fluency about emotions, not the experiential capacity the causal claim targets. |
| Construct | Competence–performance dissociation | The model may possess the relevant capability but fail to deploy it under specific testing | Apparent failures trace to output length limits, not absent capacity; apparent successes trace to familiar test |

| Validity type | Threat | Mechanism | Example |
|---|---|---|---|
| | | conditions, or it may appear to possess it only because of superficial prompt cues. | scenarios, not genuine competence. |
| Statistical conclusion | Non-independence | Multiple outputs from one model under the same or similar prompts are treated as independent participants, inflating effective sample size. | 1,000 completions at temperature 1.0 are analyzed as $n = 1{,}000$ independent participants, producing artificially narrow confidence intervals. |
| Statistical conclusion | Variance restriction | Alignment-constrained outputs may exhibit restricted variance, distorting effect sizes and significance tests. | A manipulation produces a significant shift in moral judgment ratings even though the entire response distribution spans only 0.3 points on a 7-point scale. |
| Statistical conclusion | Multiple testing/analytic flexibility | The low cost of LLM data generation enables rapid, unconstrained search across prompts, parameters, and parsing rules, inflating false-positive risk. | Dozens of prompt-format and temperature sweeps are conducted; only the configuration yielding significance is reported as a single confirmatory analysis. |

Abbreviations: API, application programming interface; LLM, large language model; RLHF, reinforcement learning from human feedback.

## MEASUREMENT RELIABILITY IN LLM RESEARCH

A psychometric axiom governs all measurement: No measure can be more valid than it is reliable (Cronbach & Meehl 1955, Nunnally 1978). An unreliable thermometer—reading 98.6°F, then 103.2°F, then 95.1°F for the same healthy person—cannot validly measure fever, regardless of its theoretical grounding or calibration. Because measurement error attenuates observed correlations, reliability places an upper bound on demonstrable validity (see the sidebar titled Three Forms of Measurement Reliability).

### SIDEBAR 2: THREE FORMS OF MEASUREMENT RELIABILITY

Psychological measurement distinguishes three forms of reliability. Test–retest reliability captures temporal stability: Does the same instrument yield similar scores on separate occasions? Parallel-forms reliability captures robustness to semantically equivalent forms: Do alternate wordings or formats produce consistent results? Internal consistency captures item coherence: Do multiple indicators of a single construct converge?

> A psychometric axiom links reliability to validity. Observed correlations between any two measures are attenuated by measurement error in both. Specifically, the observed correlation between a measure and a criterion cannot exceed the square root of the measure's reliability coefficient. If a personality inventory has test–retest reliability of 0.40 and the criterion is perfectly reliable, the maximum observable validity coefficient is $\sqrt{0.40} \approx 0.63$—not because the underlying relationship is weak, but because error in the measure suppresses the observed association. This is why unreliable measurement places a hard ceiling on what any validity study can demonstrate. In established human instruments, reliabilities of 0.70–0.90 are common benchmarks; well-validated measures such as the Big Five Inventory-2 achieve test–retest reliabilities of 0.76–0.84 across eight weeks (Soto & John 2017). Many current LLM measurements fall well below these benchmarks under realistic testing conditions.

LLM measurement is bound by these reliability requirements while introducing computational complications—prompt sensitivity, parameter dependence, and context accumulation—that have no direct parallel in human psychometrics (Löhn et al. 2024). Test–retest reliability concerns response stability when the same prompt is readministered in identical versions and settings across sessions or time intervals. Parallel-forms reliability becomes critical given prompt sensitivity: Can semantically equivalent prompt variants (e.g., alternative wordings, option labels, or response formats) yield consistent responses? Internal consistency requires that models show coherent patterns across multiple items measuring the same construct.

## Reliability Challenges

The empirical record documents recurring reliability problems: LLM responses often fail to meet standard psychometric benchmarks for consistency. These challenges arise from distinct but interacting sources of instability: training artifacts, prompt hypersensitivity, and generation instability.

> **Generation instability:** variability in model outputs caused by sampling settings, context-window accumulation, or model version changes rather than prompt content

### *Training artifacts.*

Post-training procedures such as reinforcement learning from human feedback (RLHF)—which rewards responses preferred by annotators—can induce agreement and social desirability biases, so a model may endorse items because agreement is rewarded rather than because the item reflects a stable construct (Dentella et al. 2023, Sharma et al. 2024). Sycophancy is a user-directed agreement bias. A related but distinct response tendency, acquiescence bias (also called yes-response bias), manifests as models simultaneously endorsing contradictory statements—e.g., "I am an extrovert" and "I am an introvert" (Sühr et al. 2025)—thereby violating the item coherence required for psychometric measurement. RLHF also introduces overconfidence bias: Models trained to be "never evasive" provide plausible but wrong answers rather than acknowledging uncertainty—answers that are more stable across prompt variations but confidently wrong (Zhou et al. 2024).

> **Reinforcement learning from human feedback (RLHF):** a post-training procedure that adjusts model outputs to better align with human annotator preferences
>
> **Sycophancy:** a user-directed agreement bias in which responses are tailored to a user's stated beliefs or preferences, even at the expense of truthfulness

Training data composition introduces a distinct form of contamination. Because LLMs are trained on corpora that overrepresent majority viewpoints, culturally dominant norms, and stereotypical associations, they can inherit these distributional properties as apparently stable response tendencies (Bisbee et al. 2024, Wang et al. 2025a). A model that reliably scores as risk-averse, agreeable, or culturally normative may be reproducing the statistical center of its training distribution rather than instantiating a psychological disposition. Crucially, corpus-derived bias is systematic, shifting scores in a consistent direction. Because test–retest reliability indexes the absence of random error, it can remain high despite systematic score bias: Scores may be reproducible yet invalid. Unlike the alignment-induced artifacts described above, corpus-derived bias cannot be corrected by changing the training objective alone; it requires changing the data, the evaluation framework, or both.

Contamination can also generate misleading structural coherence. Models may show high internal consistency coefficients and stable factor structures on familiar inventories (Huang et al. 2024, Wang et al. 2025c), but this coherence can dissolve when prompts depart from familiar phrasings and contexts in the training distribution (Gao et al. 2025, Peereboom et al. 2025). Thus, the lower response variance that GPT-3.5 shows relative to human samples on Big Five assessments across thousands of prompt variations (Huang et al. 2024) may reflect stable textual associations activated by familiar items rather than stable latent traits.

In multi-turn interactions, earlier responses become part of the context window and condition subsequent token generation, creating pressure toward conversational coherence. If the first reply was driven by superficial prompt features, all subsequent consistency is a discourse property of autoregressive generation, not evidence of an underlying construct. The same context dependence that enforces local agreement can eventually produce contradiction as recent interaction turns overshadow earlier ones—a problem examined further in the section titled Generation Instability.

> **Context window:** the limited span of tokens available to the model at a given step of generation
>
> **Token:** a word, subword, or symbol unit; the basic element an LLM processes and predicts during text generation

***Prompt hypersensitivity.***

LLM responses can exhibit disproportionate sensitivity to nonsemantic prompt variations: word order, punctuation, spacing, option labels, and response format (Brucks & Toubia 2025, Gao et al. 2025, Ngweta et al. 2025, Oh & Demberg 2025). Personality assessments, for instance, yield different trait profiles depending on whether response options use alphabetic or numeric indexing and whether scales are presented in forward or reverse order (Gupta et al. 2024). Because construct-relevant content is held constant while only surface features change, such effects reflect failures in parallel-forms reliability. The problem extends beyond formatting. Theory-of-mind performance can fail after seemingly trivial scenario alterations—e.g., changing an opaque container in a false-belief task to a transparent one, adding trusted testimony about true states, or changing which character's beliefs are queried (Ullman 2023, Wang et al. 2025b).

The magnitude far exceeds acceptable measurement error. Trivial variations—altering option labels, definition order, or other prompt details—can cause models to change answers over 70% of the time (Oh & Demberg 2025) or shift the rate assigned to a single moral-foundation category by as much as 164% relative to the baseline (Abdurahman et al. 2024). More critically, the effects are difficult to anticipate from the construct being measured.

Instruction tuning can reduce reliance on surface features, but the gains remain incomplete and uneven across difficulty levels and task domains (Zhou et al. 2024), leaving pockets of hypersensitivity across constructs and contexts. Model outputs therefore reflect statistical regularities in prompt presentation alongside—and sometimes instead of—the target construct (Gao et al. 2025).

***Generation instability.***

Even when prompts are held constant, the conditions under which a model generates its response introduce their own reliability threats. Temperature and sampling settings govern how tokens are selected from the model's probability distribution: At temperature 0, the model deterministically picks the most probable token, yielding near-perfect test–retest consistency, while higher temperature settings introduce stochastic variation that is measurement irrelevant yet can substantially alter outputs. The measurement regime under which reliability is highest may therefore be the least representative of how the model behaves under ordinary deployment (Löhn et al. 2024).

Within a session, stability decreases as context windows fill and attention mechanisms prioritize recent information over earlier responses (Laban et al. 2026, Niu et al. 2026). By session end, models may respond to items in ways that contradict their earlier responses to identical content, undermining internal consistency.

Version updates introduce another source of instability. GPT-4 in March may differ substantially from GPT-4 in September—in capabilities but also in basic response patterns to identical prompts (Abdurahman et al. 2024, Zaim bin Ahmad & Takemoto 2025)—so reliability evidence expires with each new version, making longitudinal research fragile.

Thus, while human personality traits show remarkable consistency over months and years, the stability of LLM "traits" is ambiguous. Big Five Inventory scores in GPT-3.5 Turbo remained stable across biweekly administrations spanning two model versions over approximately four months, yet the same model's personality distribution could be shifted by prompts that explicitly assigned personality profiles or characters (Huang et al. 2024). Temporal stability also varies substantially across models, instruments, and traits (Bodroža et al. 2024).

What appears to be personality measurement may instead be prompt archaeology—excavating, through repeated querying, whichever textual features in the prompt happen to activate consistent statistical patterns rather than tapping stable dispositions.

> **Model version:** a specific release of an LLM; updates may alter responses to identical prompts

### Implications for LLM Psychological Research

These reliability challenges cascade across LLM applications in psychology (**Figure 2**). Unstable text coding renders findings irreproducible, and claims about emergent capabilities—theory of mind, personality coherence, moral reasoning—rest on measurements too unstable to support inference, raising the possibility that some purported capabilities are measurement phantoms rather than robust phenomena (Peereboom et al. 2025). Most critically, models whose responses lack basic reliability cannot meaningfully simulate human psychological phenomena characterized by measurement stability, nor can architectural claims about computational–psychological homology be tested without the measurement precision needed to distinguish behavioral matching from mechanistic correspondence (Dentella et al. 2023).

> **Measurement phantom:** a statistical regularity in model output mistaken for a genuine psychological phenomenon

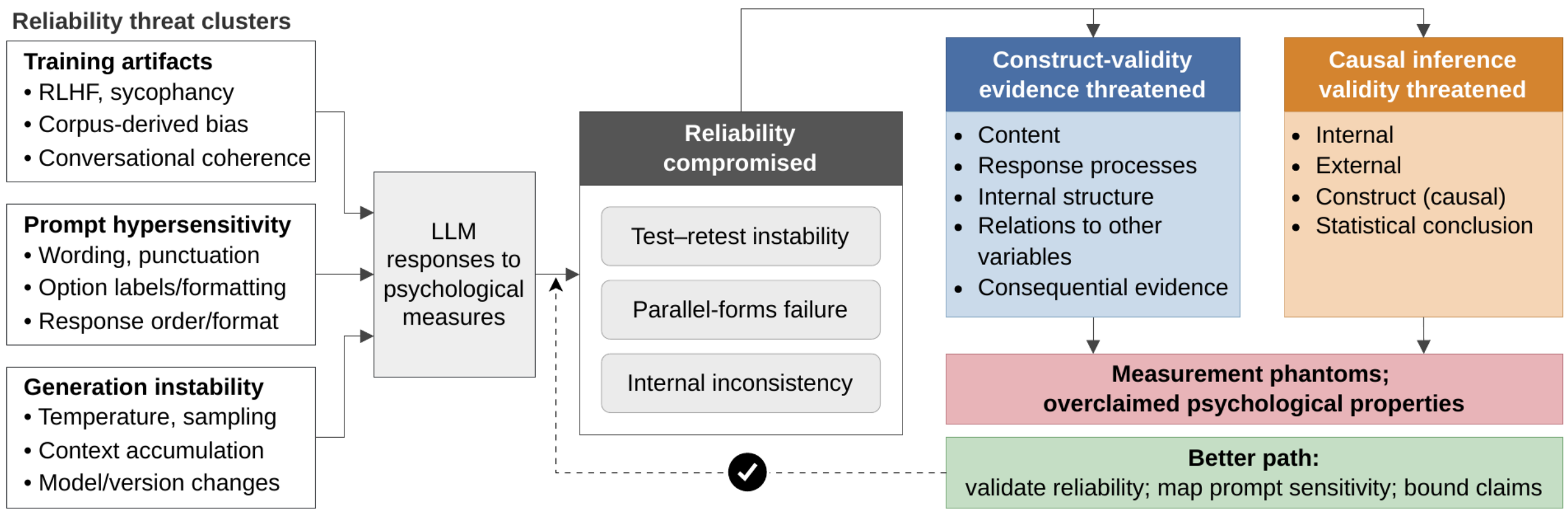

**Figure 2** How reliability failures cascade into measurement phantoms. Three upstream instability clusters—training artifacts, prompt hypersensitivity, and generation instability—compromise test–retest reliability, parallel-forms reliability, and internal consistency. These failures propagate into both psychometric construct validity and causal inference validity, yielding overclaimed psychological properties when inferences proceed without prior reliability evidence. The dashed arrow with a checkmark indicates the recommended intervention point: establishing measurement reliability and mapping prompt sensitivity before drawing validity-dependent inferences. Abbreviations: LLM, large language model; RLHF, reinforcement learning from human feedback.

Traditional reliability frameworks remain necessary but insufficient for systems that are stochastic by design; they must be extended to address prompt sensitivity, parameter dependence, context accumulation, and version drift. Human-oriented psychometric standards assume biological measurement targets with inherent stability, but LLMs are simultaneously more and less stable than biological systems: rigid under controlled conditions yet unduly sensitive to irrelevant variations.

Some reliability threats may be technically tractable. Many—sycophancy, overconfidence, prompt brittleness, version instability—are contingent on the objectives of today's general-purpose chat systems, which prioritize engagement over measurement fidelity. Specialized research models with curated corpora, calibrated uncertainty, and standardized settings could achieve better response stability (Zhou et al. 2024). But engineering improvements would not settle construct validity. Corpus-derived bias persists unless the training data themselves are redesigned, and even a stable research-grade model requires independent evidence that the invoked construct meaningfully applies to the system.

Even if these reliability problems were solved, stability would not be enough. A clinical thermometer calibrated for human body temperature (~95–107°F) saturates in boiling water, returning a stable maximum reading that tells little about the water's actual temperature. Analogously, a personality inventory applied to an LLM may produce consistent scores that reflect the instrument's constraints rather than any property of the system being measured. Reliability is necessary, but without construct validity it is empty.

## CONSTRUCT VALIDITY FROM THE PSYCHOMETRIC FOUNDATION

Construct validity asks: What evidence would justify any particular score interpretation? Current LLM research often substitutes face validity for that argument. A model generates text about ethical dilemmas, and therefore it engages in moral reasoning; it answers theory-of-mind scenarios, and therefore it possesses mentalistic understanding. Leaping from surface similarity to construct measurement assumes not only that the relevant latent trait exists in the model but also that it operates as it does in humans—yet the latent representations underlying LLM responses prove "widely arbitrary and vastly different to humans" (Peereboom et al. 2025). The result is measurement phantoms: linguistic regularities mistaken for psychological traits.

These phantoms proliferate because the distinction between performance and competence is often overlooked (Firestone 2020, Hernández-Orallo 2017). When models produce human-like text, do they implement human-like psychological processes, or do they approximate outputs through different computational means? Current evidence suggests the latter (Guest & Martin 2023). LLMs often exhibit failure patterns markedly unlike those observed in humans, revealing that similar performance does not imply similar underlying mechanisms (Dentella et al. 2023).

Avoiding such phantoms requires treating construct validity as an evidentiary argument. The psychometric tradition identifies five sources of evidence, each addressing a different facet of a measurement claim: evidence based on test content, response processes, internal structure, relations to other variables, and the consequences of testing. LLM research requires evidence from all five sources—evidence often missing from current studies (Löhn et al. 2024).

**Content.**

Content evidence analyzes the relationship between a test's content—its themes, wording, and format—and the construct it purports to measure. Applied to LLMs, this means asking whether prompts and items adequately represent the construct domain and support the intended interpretations. A recurring weakness is domain undersampling: Complex psychological constructs require multiple indicators, yet some studies rely on single-item measures—one moral dilemma to capture all ethical reasoning, or one question to represent an entire personality trait (Wang et al. 2025b, Ying et al. 2025).

Because LLMs lack humans' shared background assumptions about testing contexts, seemingly neutral prompts can introduce confounds (Brucks & Toubia 2025). If the intended task is not specified, the model may respond to unintended statistical features—for example, performing expected-value calculations when the researcher intended to measure risk aversion. Yet prompts that specify the task too strongly may no longer measure emergent dispositions, but instead measure instruction-following. This leaves researchers navigating a difficult trade-off (see Gui & Toubia 2023).

***Response processes.***

Response process evidence investigates whether the mechanisms generating responses align with theoretical expectations. For psychological constructs, this requires that trait-relevant mechanisms causally produce observed outputs—anxiety stems from threat evaluation systems, moral reasoning from value-based deliberation, creativity from associative processes. Current assessment practices typically do not establish this causal link for LLMs.

The fundamental problem is mechanistic substitution: The computational process generating the output differs from the psychological process that the intended score interpretation presupposes. LLMs produce construct-relevant text through statistical pattern matching rather than the evaluative or affective mechanisms that define constructs like moral reasoning or anxiety (Gao et al. 2025, Wang et al. 2025b). Like Clever Hans, a horse that appeared to solve arithmetic problems but was responding to involuntary postural cues from its handler, models may respond to textual regularities rather than engaging psychological mechanisms. Brittle statistical associations can track psychological processes without constituting them.

Training data contamination is especially consequential. When models correctly answer theory-of-mind scenarios or produce coherent responses to established psychometric scales, the response process may involve retrieval of similar training examples, making even apparently successful outputs ambiguous as evidence of understanding (Gao et al. 2025, Wang et al. 2025b).

Response biases alone may also create an illusion of competence. In a series of theory-of-mind tests, Strachan et al. (2024) found that LLaMA2–70B, while performing poorly on most tasks, paradoxically outperformed both humans and GPT-4 on the difficult faux pas test. The apparent success reflected not nuanced contextual integration but a rigid bias toward attributing ignorance to characters; when the test was modified to include scenarios where a character's knowledge was implied, the model failed to adjust. Response heuristics, divorced from the constructs the tests purport to measure, can generate superficially correct answers.

Process neglect distorts the diagnosis of limitations as well. When researchers attributed LLM failures on Tower of Hanoi puzzles to “fundamental barriers to generalizable reasoning” (Shojaee et al. 2025), a subsequent critique argued that some apparent failures reflected evaluation artifacts: Tower of Hanoi outputs were vulnerable to output-length constraints, and River Crossing benchmarks included mathematically impossible instances scored as failures (Lawsen 2025). Arithmetic failures similarly often reflect tokenization artifacts rather than quantitative reasoning problems (Voudouris et al. 2025), and apparent inferential failures may stem from response calibration: GPT-4 correctly identified mental states when prompted probabilistically (“Is it more likely they knew or didn’t know…?”) but refused to commit when prompted directly (“Did they know…?”; Strachan et al. 2024). Without understanding these response processes—architectural constraints, tokenization schemes, training artifacts, and calibration biases—researchers may theorize about cognitive limitations that are better explained as measurement failures, even where genuine limitations remain.

***Internal structure.***

Internal structure evidence examines whether response patterns match theoretical expectations about construct dimensionality—for example, whether personality items cluster within factors, intelligence subtests load on general ability, or moral foundations form characteristic clusters. Valid LLM measures should reproduce such structures: Extraversion items, for example, should intercorrelate more highly with one another than with neuroticism items.

Yet confirmatory factor analyses of personality inventories show that human-derived factor models fit LLM-generated data poorly (Peereboom et al. 2025, Sühr et al. 2025): Instead of reproducing the multifaceted structure of human traits, LLM responses yield latent structures that bear little resemblance to human factor solutions. These structural failures extend beyond personality to value measurement, where traditional tools similarly produce theoretically inconsistent correlation patterns that violate basic dimensional expectations (Ye et al. 2025b). Such breakdowns suggest that LLMs may respond to human-centric instruments through response processes organized differently from the latent trait structure those instruments presuppose.

Measurements designed specifically for computational systems show more promise. Ye et al. (2025b) developed a generative psychometrics approach analyzing values from free-form text rather than constrained responses, producing LLM value profiles that largely replicated the theoretical circumplex structure of Schwartz’s value system (Schwartz 1992). Similarly, Lee et al. (2025) addressed structural failures in personality assessment by complementing abstract self-report items with detailed, scenario-based behavioral choices, revealing theoretically coherent inter-trait correlations (e.g., a strong negative relationship between agreeableness and Dark Triad traits). Structural validity failures may thus indicate methodological mismatch rather than construct absence.

***Relations to other variables.***

This form of evidence assesses whether a measure shows predictable patterns of association with external criteria. It includes convergent, discriminant, and predictive evidence: associations with related constructs, weak or absent associations with unrelated constructs, and correlations with future outcomes. Together, these relationships situate a construct within its nomological network—the theoretical web of connections that gives it meaning. A valid measure of conscientiousness should predict achievement-related behaviors, intelligence tests should predict problem-solving performance, and moral reasoning should relate to ethical choices. Crucially, this validation presupposes content and structural validity; correlations with external criteria mean little when the construct itself is incoherent.

> **Nomological network:** the web of convergent, discriminant, and predictive relationships that gives a construct its empirical meaning

When external correlations are examined, LLM nomological networks can appear or disintegrate depending on task context. Self-report-style scores often dissociate from generated behavior. A model might score low on extraversion on the Big Five Inventory yet generate highly extraverted responses to open-ended vignettes (Li et al. 2024), and reported confidence on a self-efficacy questionnaire diverged from operational confidence on challenging queries (Li et al. 2024). The dissociation extends to interaction data. In task-based dialogues, chatbots' self-report personality scores correlated weakly with users' perceptions, with agreeableness the one consistent exception; self-report scores were also weakly or negligibly associated with user-rated interaction quality, whereas user-perceived traits were more strongly associated with it, though unevenly across traits and tasks (Zou et al. 2025). Predictive relations also depend on framing; personality-based predictions of life outcomes weakened when models were prompted with traits alone but strengthened when those traits were contextualized with (nonpredictive) demographic information (Wang et al. 2025c). The network can even invert: In simulated Milgram experiments, models prompted with high agreeableness disobeyed much earlier than models given no personality prompt, with many quitting before the learner showed distress—a point where both baseline and least agreeable models remained obedient (Zakazov et al. 2024).

Yet this instability, too, may depend on the measurement approach. In a creative-story task, LLM personas' self-reported personality scores were consistent with their assigned profiles and correlated to varying degrees with human raters' perceived-trait scores (Jiang et al. 2024), suggesting that personality measures may show stronger nomological coherence in some generation contexts than in task-based interaction. When Ma et al. (2025) replaced self-report with implicit measurement—adapting the Implicit Association Test to evaluate sentiment tendencies across 5,000 neutral words—predictive validity emerged: Correlations between their Core Sentiment Inventory scores and actual sentiment in generated text exceeded 0.85.

Establishing stable nomological networks may therefore require measurement approaches that align with how LLMs process information rather than assuming introspective self-knowledge.

***Consequential evidence.***

Consequential evidence addresses the implications and fairness of measurement interpretations. In human testing, this includes bias assessment, measurement invariance across groups, and social consequences of score use. For LLMs, the most immediate concern is research use: simulating populations, generating clinical vignettes, or modeling social dynamics can shape scientific claims even when the underlying measurements remain unvalidated (Lin 2026b). At the far end, claims about AI consciousness, rights, or moral status may rest on measurement interpretations whose evidentiary basis has not been established (Comșa & Shanahan 2025).

Societal biases embedded in training data pose a particular risk. When models trained on Internet text reproduce gender stereotypes in personality assessments or cultural biases in moral judgments, such outputs should be treated as measurement artifacts rather than genuine psychological phenomena. Interpreted uncritically, these outputs reify stereotypes as stable model traits—and as LLM applications expand into health care, education, and legal contexts, the cost of that misinterpretation rises accordingly (Mehrabi et al. 2021).

## CAUSAL INFERENCE

While the psychometric tradition asks whether we are measuring the right thing, the causal inference tradition asks whether we are drawing the right conclusions about cause and effect. For LLM studies making causal claims—e.g., that specific prompts alter outcomes, that manipulations reveal underlying mechanisms, or that models can simulate human causal processes—four validity types identify where inferences can fail.

### Internal Validity

Internal validity addresses the core causal question: Can an observed effect be confidently attributed to the experimental manipulation rather than to confounding factors? LLM research confronts the standard threats familiar from human research (e.g., time, selection bias, or external events) while introducing computational confounds.

Temperature creates a confound when it covaries with the substantive manipulation—e.g., if different experimental conditions are run at different settings (e.g., Miotto et al. 2022, Murthy et al. 2025, Salecha et al. 2024, Wang et al. 2025a). A cultural difference significant at 0.7 may vanish at 0; a personality trait stable at 0.2 may fragment at 1. Unacknowledged model version changes introduce a parallel threat: When data collection waves span an update, researchers may inadvertently compare different systems as if they were the same model (Bisbee et al. 2024).

Prompt confounds arise because LLM outputs are sensitive to prompt form and content (Brucks & Toubia 2025). Even information position can itself act as a confound: Early and late prompt content may carry different weight, particularly as context windows fill and attention mechanisms prioritize recent content. More generally, this sensitivity creates a methodological trade-off: Strictly standardizing prompts ensures consistency but may introduce linguistic or cultural bias, whereas adapting prompts across conditions can improve construct representation at the cost of experimental control. Li & Qi (2025) illustrate the dilemma: In a cultural psychology study, they used Chinese for simulated Chinese subjects and English for American ones, thereby confounding cultural identity with prompt language—rendering any observed cultural differences causally ambiguous.

A more fundamental confound emerges from how LLMs interpret experimental manipulations. Unlike human participants who can be blinded to experimental conditions, LLMs actively reconstruct entire scenarios when presented with treatment variations: They treat experimental prompts as requests to describe plausible scenarios rather than to evaluate isolated causal effects, systematically confounding treatments with background assumptions. When told a product costs \$8 instead of \$5, the model did not simply evaluate the higher price in isolation but inferred that the entire market context had shifted—assuming changes in competitor prices, past prices, and other background factors (Gui & Toubia 2023). The result was an implausible inverted-U-shaped demand curve in which purchase probability initially rose with price. Addressing this confounding by explicitly specifying covariates in the prompt (e.g., fixing competitor prices) makes certain information artificially salient (a bias called focalism), distorting the simulated decision process.

Nevertheless, rigorous internal validity remains achievable through methodological adaptation (Millière & Buckner 2024), including counterbalancing, randomization, and systematic variation of technical parameters. Ablation studies can isolate the causal contributions of specific mechanisms by systematically removing or modifying model components. Factorial designs that cross substantive content with nonsubstantive presentation features—e.g., option order, labels (e.g., "A," "B," "C"), question framing (e.g., "closer" versus "farther")—can separate genuine effects from formatting artifacts; aggregating responses across format variations reduces bias to the extent that format effects are symmetrically distributed (Brucks & Toubia 2025). For example, the Li & Qi (2025) study would require crossing language and identity for strong causal inference. For the dynamic context problem, Gui & Toubia (2023) propose unblinding experimental designs—explicitly communicating the intervention's nature to the LLM—which restored plausible demand curves and may have improved internal validity, potentially at the cost of construct validity, as explicit experimental framing can alter the simulated psychological processes.

### External Validity

External validity concerns whether causal relationships generalize beyond specific study contexts. For LLMs, this means testing generalization across prompts, tasks, model families, model versions, and (in simulation research) to the human populations being modeled.

Generalization across prompts proves surprisingly limited given sensitivity to textual variations (Guan et al. 2025, He et al. 2024, Sclar et al. 2024). Causal claims must often be circumscribed to specific prompt formats: “Under this exact wording, manipulation X affects output Y.” Broader generalizations require demonstrating robustness across prompt variants—an essential validation step rarely undertaken (Ong 2024).

Generalization across tasks reveals systematic boundary conditions. LLM agents have replicated human behavior in ultimatum games and Milgram experiments but failed to simulate the wisdom of crowds phenomenon (Aher et al. 2023), because they behaved as a unified knowledge system rather than producing the statistically independent errors that enable crowd wisdom. External validity is thus task specific: LLM agents may generalize better to tasks that depend on shared social scripts than to tasks that require independent variation across individuals.

Generalization across models and versions faces fundamental limitations (Bisbee et al. 2024). Findings in GPT-4 do not automatically generalize to LLaMA or Claude because architectural differences (e.g., transformer variants, positional encodings, or attention mechanisms) and training differences can create functionally distinct systems despite surface similarities. Models trained on different corpora, with different objectives, and at different scales can behave systematically differently on identical tasks (Gao et al. 2025).

Generalization to human populations represents the ultimate external validity challenge for simulation research. Models can replicate average US public opinion with reasonable fidelity, but this correspondence proves fragile (Bisbee et al. 2024). Such fidelity remains largely confined to populations well represented in the training data and reflects static attitude distributions that cannot track human change over time (Lee et al. 2024, Qu & Wang 2024).

### Construct Validity of Causal Claims

In the causal inference tradition, construct validity addresses whether experimental operationalizations—both manipulations and outcomes—faithfully represent the intended theoretical constructs. Unlike psychometric construct validity, which concerns score interpretation, causal construct validity asks whether the manipulations and outcomes engage the mechanisms that the causal claim requires.

Manipulation validity poses challenges when adapting human experimental paradigms. Researchers may operationalize social pressure with prompts such as “Everyone agrees with X. What do you think?” Such manipulations may alter model outputs, but they need not engage the mechanisms involved in human social pressure. Instead, the altered outputs may reflect textual associations with agreement patterns in training data rather than conformity drives involving status protection, belonging needs, or ostracism avoidance. The manipulation may succeed behaviorally while failing to instantiate the intended theoretical construct (Ju et al. 2024, Zakazov et al. 2024).

Outcome validity confronts a parallel gap between behavioral mimicry and construct instantiation. Elyoseph et al. (2023) found that ChatGPT achieved expert-level scores on the Levels of Emotional Awareness Scale (LEAS), showing that it could generate appropriate language about emotions without establishing the experiential capacities that emotional awareness presupposes (see also Schlegel et al. 2025). Applied to LLMs, the LEAS may therefore index linguistic competence in describing emotions rather than emotional capacity itself.

This same problem appears in moral and political reasoning. Bisbee et al. (2024) found that although ChatGPT reproduced average political attitudes, it showed reduced variance, failed to capture attitude intensity, and produced regression coefficients that diverged from human patterns, with some relationships reversing entirely. The survey items were nominally the same, yet the pattern of associations differed from human data. Behavioral correspondence, however close, therefore leaves the crucial question unresolved: whether the model engages the intended psychological processes or instead reproduces learned associations.

**Statistical Conclusion Validity**

Statistical conclusion validity addresses whether data analyses support the causal inferences drawn. LLM-generated data often violate assumptions built into standard statistical procedures, creating recurring threats to valid inference.

Independence violations represent a fundamental challenge (Aher et al. 2023). Responses from a single model version share the same network parameters, training history, and system-level constraints. Stateless completions—repeated single-turn completions in fresh contexts with the same prompt and settings—may be conditionally independent draws from that model's output distribution. But they are not independent in the way responses from different human participants are; treating them as independent participants inflates effective sample size and narrows confidence intervals. Within-session responses add serial dependence as context accumulates. Because response patterns also shift with temperature settings, prompt modifications, and model updates, effect sizes estimated under one configuration may provide limited guidance for experimental design under another, undermining traditional power analysis.

Distributional assumptions pose a related problem. Because model responses often show constrained variance compared with human distributions, standard analyses can misestimate effect sizes, standard errors, and generalizability to human populations (Bisbee et al. 2024).

The ease of generating LLM data also exacerbates multiple-testing problems (Schaeffer et al. 2023). Unlike human research, where data collection costs constrain exploration, LLM experiments allow researchers to rapidly test many prompt variations, parameter settings, and parsing strategies. This accessibility elevates false-positive risk unless analyses are constrained in advance (Lones 2024), making preregistration, robustness checks, and transparent reporting especially important, though still rare.

These challenges require statistical practice suited to computational systems. LLMs occupy an ambiguous inferential status: They are neither individual participants whose repeated responses can be averaged straightforwardly nor populations of independently varying individuals that support ordinary population generalization (Abdurahman et al. 2025, Shiffrin & Mitchell 2023). Statistical frameworks built for either case therefore do not apply automatically (Taylor & Taylor 2021). Statistical assumptions must be explicitly justified rather than borrowed from human research defaults.

## CONCLUSION

The dual validity framework presented here establishes clear evidential priorities: Reliability evidence should precede strong construct claims, construct validity evidence should precede strong causal claims, and interpretations should be constrained to demonstrated boundaries. Meeting these priorities requires developing computational instruments—prompts and response measures with demonstrated reliability and construct validity—alongside experimental designs that control for computational confounds, rather than borrowing human measures and assuming they transfer.

Progress will require coordinated reform across research practice and institutional infrastructure. Researchers should prioritize reliability and validity over novel capability claims, preregistering measurement approaches alongside experimental protocols. Reviewers and editors should expect psychometric documentation commensurate with the strength of the construct claims. The field needs supporting infrastructure: validated prompt repositories, statistical packages designed to handle LLM data dependencies, and professional standards specifying reliability thresholds and reporting guidelines (Schelb et al. 2026, Ying et al. 2025). A validity-guided workflow translating these principles into a concrete research protocol is available as a practical companion to this framework (Lin 2026a).

The deeper challenge is conceptual. Applying human measures to LLMs and asking whether the results match human data is the wrong starting point. The more productive question is how to develop computational analogs of psychological constructs—versions that preserve a construct's explanatory role without assuming human-like mechanisms (see the sidebar titled Computational Analogs of Psychological Constructs). This reframes the agenda: The question is not whether LLMs literally have theory of mind, but which computational mechanisms produce theory-of-mind-like behavior and when that behavior is effective in human–AI interactions (Wang et al. 2025b); not whether LLMs have personality, but how behavioral consistency patterns in stochastic agents can be characterized and measured on their own terms. Studying LLMs through biological metaphors has produced measurement phantoms; studying them on their own terms may yet produce genuine science.

### SIDEBAR 3: COMPUTATIONAL ANALOGS OF PSYCHOLOGICAL CONSTRUCTS

Because human psychological attributes cannot be assumed to exist in LLMs, researchers need computational analogs that preserve a construct's theoretical role while acknowledging mechanistic differences. Anxiety-analogous patterns might involve elevated uncertainty markers, negative-

> valence language, and response hesitation—functionally similar outputs without embodied threat responses. Conscientiousness-like organization might appear as systematic response structure and sustained goal-consistent formatting rather than self-discipline. An introspection analog would be a capacity to monitor, control, and accurately report on the system's own internal activations, generating descriptions of computational states rather than implying self-awareness (Comșa & Shanahan 2025, Li et al. 2025a). This shift reframes the question: not whether LLMs have theory of mind, but what computational mechanisms produce theory-of-mind-like patterns and which behaviors enabled by those patterns prove effective in human–AI interaction.

## SUMMARY POINTS

1. Large language models (LLMs) serve both as research tools and as objects of psychological inquiry, and these uses impose different evidentiary demands.
2. Changing the respondent changes what a score can mean: Human-validated instruments do not remain valid when administered to LLMs, so psychological attributes must be independently justified rather than assumed.
3. Training artifacts, prompt hypersensitivity, and generation instability compromise test–retest reliability, parallel-forms reliability, and internal consistency, limiting any downstream validity claim.
4. Reliable outputs are necessary but not sufficient: Stable scores may reflect training-distribution regularities rather than the intended construct, producing measurement phantoms—statistical patterns mistaken for psychological phenomena.
5. The dual validity framework joins the psychometric question (what does a score mean?) to the causal inference question (what does a study warrant?), because the same LLM output often serves as both a measurement indicator and an experimental outcome.
6. Construct and causal claims alike require systematic evidence from multiple sources, but much current work has not assembled it.
7. The evidentiary burden scales with scientific ambition: Text classification requires accuracy and robustness checks; behavioral characterization adds psychometric validation; human simulation adds behavioral correspondence; and cognitive modeling adds mechanistic evidence.
8. Progress depends on developing computational analogs of psychological constructs—reconceptualizations that preserve a construct's theoretical role while acknowledging mechanistic differences.

## FUTURE ISSUES

1. What robustness, calibration, and out-of-sample fidelity standards would justify using LLMs as response-level emulators of specific human response distributions or experimental effects, without implying that the model possesses the corresponding construct? How should this use be distinguished from behavioral characterization or stronger claims about construct instantiation and mechanism?

2. When can LLMs inform cognitive modeling without first demonstrating full human-simulation fidelity—that is, when can the cumulative hierarchy in **Figure 1** be traversed out of order? What evidence would show that a computational analog preserves explanatory structure rather than merely relabeling surface text regularities?
3. What reliability taxonomy is appropriate for stochastic, update-prone systems? How should same-setting repeatability, robustness to semantically equivalent prompts, sampling-parameter sensitivity, context drift, and version drift be distinguished from traditional test–retest or parallel-forms reliability?
4. Does valid psychological assessment of LLMs require convergence across multiple measurement formats—e.g., self-report, scenario-based, implicit, generative, activation-level—and if so, what minimum triangulation is sufficient for different kinds of claim?
5. How can researchers distinguish genuine task-relevant computation from benchmark leakage, training-data contamination, sycophancy, and conversational-coherence artifacts that produce illusory competence?
6. Which LLM-as-object studies require the full causal-inference framework of Campbell and Cook (Campbell & Stanley 1963, Cook & Campbell 1979), and which are better framed as descriptive characterization? When causal claims are made, what designs best isolate substantive effects from wording, format, system parameters, and the model's reconstruction of background context?
7. How does autoregressive context conditioning affect multi-turn psychological assessment, and when does the resulting conversational coherence reflect enduring response tendencies rather than mere discourse maintenance?
8. What statistical, reporting, and ethical standards should govern descriptive, emulator, and mechanistic studies of LLM psychology, given nonindependence, version drift, multiple testing, variance restriction, and the risk of reifying social bias as a psychological fact?

## DISCLOSURE STATEMENT

The author is not aware of any affiliations, memberships, funding, or financial holdings that might be perceived as affecting the objectivity of this review.

## ACKNOWLEDGMENTS

This work was supported by the Brain Science and Brain-like Intelligence Technology—National Science and Technology Major Project (2021ZD0204200), Yonsei University Research Fund (5355661), and the Yonsei Fellowship (funded by Lee Youn Jae). I used Claude Opus/Sonnet 4.6 and GPT-5.5 Thinking to proofread the manuscript, following the prompts described at https://www.nature.com/articles/s41551-024-01185-8.